\documentclass[10pt,twocolumn,twoside]{IEEEtran}
\usepackage[utf8]{inputenc}
\usepackage{acronym,amsfonts,amsmath,amssymb,authblk,cite,graphicx,multirow}

\begin{document}

\newacro{AV}{Arousal-Valence}
\newacro{CNN}{Convolutional Neural Network}
\newacro{CCA}{Canonical Correlation Analysis}
\newacro{DASM}{Differential Asymmetry}
\newacro{DC}{direct current}
\newacro{DCCA}{Deep Canonical Correlation Analysis}
\newacro{DL}{Deep Learning}
\newacro{EEG}{Electroencephalogram}
\newacro{EMG}{Electromyogram}
\newacro{EOG}{Electrooculogram}
\newacro{LDA}{Latent Dirichlet Allocation}
\newacro{LDS}{Linear Dynamic System}
\newacro{LSTM}{Long Short-Term Memory}
\newacro{MER}{Music Emotion Recognition}
\newacro{MIR}{Music Information Retrieval}
\newacro{ML}{Machine Learning}
\newacro{MRR}{Mean Reciprocal Rank}
\newacro{NN}{Neural Network}
\newacro{pp}{percentage points}
\newacro{RNN}{Recurrent Neural Network}
\newacro{SVM}{Support Vector Machine}
\newacro{WAR}{Wavelet Artifact Removal}
\newacro{WSD}{Wavelet Semblance Denoising}

\title{Towards Deep Modeling of Music Semantics using \acs{EEG} Regularizers}
\author{Francisco Raposo, David Martins de Matos, Ricardo Ribeiro, Suhua Tang, Yi Yu}

\maketitle

\begin{abstract}
Modeling of music audio semantics has been previously tackled through learning of mappings from audio data to high-level tags or latent unsupervised spaces. The resulting semantic spaces are theoretically limited, either because the chosen high-level tags do not cover all of music semantics or because audio data itself is not enough to determine music semantics. In this paper, we propose a generic framework for semantics modeling that focuses on the perception of the listener, through \acs{EEG} data, in addition to audio data. We implement this framework using a novel end-to-end 2-view \ac{NN} architecture and a \ac{DCCA} loss function that forces the semantic embedding spaces of both views to be maximally correlated. We also detail how the \acs{EEG} dataset was collected and use it to train our proposed model. We evaluate the learned semantic space in a transfer learning context, by using it as an audio feature extractor in an independent dataset and proxy task: music audio-lyrics cross-modal retrieval. We show that our embedding model outperforms Spotify features and performs comparably to a state-of-the-art embedding model that was trained on 700 times more data. We further discuss improvements to the model that are likely to improve its performance.
\end{abstract}

\section{Introduction}

Recent advances in \ac{ML} have paved the way for implementing systems that compute compact and fixed-size embeddings of music data \cite{hu2014,wang2015,choi2017,lee2017a,lee2017b,park2017}. The design of these systems is usually motivated by the pursuit of automatic inference of music semantics from audio, by describing it in a learned semantic space. However, most of these systems are limited to the availability of labeled datasets and, more importantly, are limited to learning patterns in data solely from the artifacts themselves, i.e., solely from fixed (objective) descriptions of the object of the subjective experience. Although audio content is important and, to a certain extent, empirically proven to be effective in representing music semantics, it does not account for all factors involved in music cognition. Therefore, since music is ultimately in the mind, understanding the process of its perception by focusing on the listener is necessary to effectively model music semantics \cite{widmer2016}.

In order to address the lack of attention to the listener in previous \ac{MIR} approaches to music semantics, we focus on the neural firing patterns that are manifested by the human brain during perception of music artifacts. These patterns can be recorded using \ac{EEG} technology and effectively employed to study music semantics. Previous research has applied \acp{EEG} for studying the correlations between neural activity and music, yielding important insights, namely, regarding appropriate electrode positions and spectrum frequency bands \cite{schmidt2001,altenmuller2002,sammler2007,lin2010,duan2012,hadjidimitriou2012,daly2014,thammasan2014,thammasan2016}.

We present a generic framework to model multimedia semantics. We leverage multi-view models, that learn a space of shared embeddings between \acp{EEG} and the chosen medium, as an implementation. We instantiate this framework in the context of music semantics, by proposing a novel end-to-end \ac{NN} architecture for processing audio and \acp{EEG}, making use of the \ac{DCCA} loss objective. The learned space is capable of capturing the semantics of music audio by using subjective \ac{EEG} signals as regularizers during its training. In this sense, the framework defines music semantics as a by-product of the interplay between audio artifacts and perception of listeners, being only theoretically limited by the measuring precision of the \acp{EEG}. We evaluate the effectiveness of this model in a transfer learning setting, using it as a feature extractor in a proxy task: music audio-lyrics cross-modal retrieval. We show that the proposed framework is able to achieve very promising results when compared against standard features and a state-of-the-art model, using much less data during training. We also discuss improvements to this specific instance of the framework that can improve its performance.

This paper is organized as follows: Sections \ref{sec:audio} and \ref{sec:eeg} review related work on modeling audio semantics and \ac{EEG}-based \ac{MIR}, respectively; Section \ref{sec:dcca} introduces \ac{DCCA} and Section \ref{sec:architecture} proposes our novel \ac{NN} architecture for modeling audio and \ac{EEG} correlations; Section \ref{sec:dataset} explains the \ac{EEG} data collection processs; Section \ref{sec:experiments} details the experimental setup; Section \ref{sec:results} presents and discusses results as well as the advantages of this approach to modeling music semantics; and Section \ref{sec:conclusions} draws conclusions and proposes future work.

\section{Music Audio Semantics}
\label{sec:audio}

Several proposed approaches can be used for modeling music by estimating an audio latent space. Gaussian-\ac{LDA} \cite{hu2014}, proposed as a continuous data extension of \ac{LDA} \cite{blei2003}, has been successfully applied in an audio classification scenario. This unsupervised approach estimates a mixture of latent Gaussian topics, that are shared among a collection of documents, to describe each document. Even though this approach requires no labeling, it is yet to be proven to be able to infer robust music features. Music audio has also been modeled with Gaussian mixtures in the context of \ac{MER} \cite{wang2015}, where the affective content of music is described by a probability distribution in the continuous space of the \ac{AV} plane \cite{russell1980,thayer1989}. This probabilistic approach is motivated by the fact that emotion is subjective in nature. However, this study only focuses on prediction of affective content and requires expensive annotation data. In order to overcome the issue of expensive data annotation, a \ac{CNN} was trained using only artist labels \cite{park2017}, which are usually available and require no annotation. This system was shown to produce robust features in transfer learning contexts. However, even though the assumption that artist information guides the learning of a meaningful semantic space is usually valid, it is not powerful enough since it breaks down in the presence of polyvalent musicians. Even when using expensive labeling, such as in \cite{choi2017}, where ``semantic'' tags were used to learn the semantic space, there are still problems such as the granularity and abstraction level of the tags not being consistent or aligned with the corresponding audio that is responsible for the presence of those tags. Heuristic attempts to solve the problem of granularity and abstraction level were proposed in \cite{lee2017b}, where several models are trained, each operating on a different time-scale, and the final embeddings consist of an aggregation of embeddings from all models. However, the label alignment issue is still unresolved, the feature aggregation step is far from optimal and it is virtually impossible to find and cover every appropriate time-scale.

Our framework differs from these related works which suffer from the previously mentioned drawbacks. As opposed to relying on explicit labels, we rely on measurements of the perception of listeners. We can think of this paradigm as automatic and direct ``labeling'' by the brain, bypassing faulty conscious labeling decisions and the ``tyranny'' of words or categories. Thus, we no longer have the labeling taxonomy issue of chosing between too coarse or too granular categories which lead to not rich enough or ambiguous categories, respectively \cite{posner2005,yang2011}. We also do not need to resort to dimensional models of emotion and, thus, to specify which psychological dimensions are worth modeling \cite{russell1980,thayer1989}. Furthermore, since both audio and \ac{EEG} signals unfold in time, we have a natural and precise time alignment between both and, thus, a more fine-grained and reliable ``annotation'' of music audio.

\section{\ac{EEG}-based \ac{MIR}}
\label{sec:eeg}

The link between brain signals and music perception has been previously explored in \ac{MER} using \ac{EEG} data. Several studies reduce this problem to finding correlations between music emotion annotations and the time-frequency representation of the \acp{EEG} in five frequency bands (in Hz): $\delta$ ($<4$), $\theta$ ($\ge4$ and $<8$), $\alpha$ ($\ge8$ and $<14$), $\beta$ ($\ge14$ and $<32$), and $\gamma$ ($\ge32$).

In \cite{thammasan2014}, 3 subjects annotated 6 clips on a 2D emotion space and had their 12-channel \acp{EEG} recorded. \ac{SVM} classification achieved accuracies of 90\% and 86\% for arousal and valence, respectively (binary classification). In \cite{altenmuller2002}, 12-channel \acp{EEG} were recorded from 16 subjects and 160 clips, revealing correlations between lateralised and bilateralised patterns with positive and negative emotions, respectively. In \cite{duan2012}, 62-channel \ac{LDS}-smoothed \ac{DASM} features extracted from 5 subjects and 16 tracks were able to achieve 82\% classification accuracy. In \cite{daly2014}, pre-frontal and parietal cortices were correlated with emotion distinction in an experiment involving 31 subjects and 110 excerpts, using 19-channel \acp{EEG}. 82\% accuracy was achieved in 4-way classification with 32-channel \ac{DASM} features extracted from 26 subjects and 16 clips in \cite{lin2010}. Correlations were also found between mid-frontal activation and dissonant music excerpts in the context of an 18 subjects and 10 clips 24-channel \ac{EEG} experiment in \cite{sammler2007}. In \cite{schmidt2001}, 59 subjects listening to 4 excerpts provided the 4-channel \ac{EEG} data which revealed that asymmetrical frontal activation and overall frontal activation are correlated with valence and arousal perception, respectively. 14-channel \acp{EEG} extracted from 9 subjects that listened to 75 clips showed correlations with emotion recognition in the frontal cortex in \cite{hadjidimitriou2012}. Binary emotion classification over time was performed in \cite{thammasan2016}, where an average 82.8\% and 87.2\% accuracy were achieved for arousal and valence, respectively.

Not all studies report the same correlations nor used the same experimental setup, but common and relevant conclusions can be found regarding features and electrode locations relevant for music perception. Power density, in the frontal and parietal regions, has been observed to correlate with emotion detection in music \cite{schmidt2001,altenmuller2002,sammler2007,lin2010,duan2012,hadjidimitriou2012,daly2014,thammasan2014,thammasan2016}. Asymmetrical power density in the frontal region was linked to music valence perception \cite{schmidt2001,altenmuller2002,daly2014,thammasan2014,thammasan2016}. A link has also been revealed between overall frontal activity and music arousal perception \cite{schmidt2001}.

In our work, we follow previously mentioned major conclusions regarding electrode positioning but not frequency bands, since our proposed architecture is end-to-end, thereby bypassing handcrafted feature selection. Furthermore, the focus of this paper is on using \ac{EEG} responses as regularizers in the estimation of a generic semantic audio embeddings space, as opposed to using \acp{EEG} for studying specific aspects of music. Note that these previous works build systems that can predict these aspects (emotion), given new \ac{EEG} input. Our approach is able to predict generic semantic embeddings given new audio input, as it needs \ac{EEG} data only during training.

\section{Deep Canonical Correlation Analysis}
\label{sec:dcca}

\ac{DCCA} \cite{andrew2013} is a model that learns maximally correlated embeddings between two views of data and is effective at estimating a music audio semantic space by leveraging \ac{EEG} data from several regularizer human subjects. It is a non-linear extension of \ac{CCA} \cite{hotelling1936} and has previously been applied to learn a correlated space in music between audio and lyrics views in order to perform cross-modal retrieval \cite{yu2017}. It jointly learns non-linear mappings and canonical weights for each view:
\begin{small}
\begin{equation}
\left(w_x^*,w_y^*,\varphi_x^*,\varphi_y^*\right)=\underset{\left(w_x,w_y,\varphi_x,\varphi_y\right)}{\operatorname{argmax}}\operatorname{corr}\left(w_x^{\bf{T}}\varphi_x\left(\bf{x}\right),w_y^{\bf{T}}\varphi_y\left(\bf{y}\right)\right)
\end{equation}
\end{small}
where $\bf{x}\in{\rm I\!R}^m$ and $\bf{y}\in{\rm I\!R}^n$ are the zero-mean observations for each view, with covariances $C_{xx}$ and $C_{yy}$, respectively, and cross-covariance $C_{xy}$. $\varphi_x$ and $\varphi_y$ are non-linear mappings for each view, and $w_x$ and $w_y$ are the canonical weights for each view. We use backpropopagation and minimize:
\begin{equation}
-\sqrt{\operatorname{tr}\left(\left(C_{XX}^{-1/2}C_{XY}C_{YY}^{-1/2}\right)^{\bf{T}}\left(C_{XX}^{-1/2}C_{XY}C_{YY}^{-1/2}\right)\right)}\
\end{equation}
\begin{equation}
C_{XX}^{-1/2}=Q_{XX}\Lambda_{XX}^{-1/2} Q_{XX}^{\bf{T}}
\end{equation}
where $X$ and $Y$ are non-linear projections for each view. $C_{XX}$ and $C_{YY}$ are the regularized, zero-centered covariances while $C_{XY}$ is the zero-centered cross-covariance. $Q_{XX}$ are the eigenvectors of $C_{XX}$ and $\Lambda_{XX}$ are the eigenvalues of $C_{XX}$. $C_{YY}^{-1/2}$ can be computed analogously. We finish training by computing a forward pass with the training data and fitting a linear \ac{CCA} model on those non-linear mappings. The canonical components of these deep non-linear mappings implement our semantic embeddings space.

\section{Neural Network Architecture}
\label{sec:architecture}

\begin{figure}[htbp]
\begin{center}
\includegraphics[width=\columnwidth]{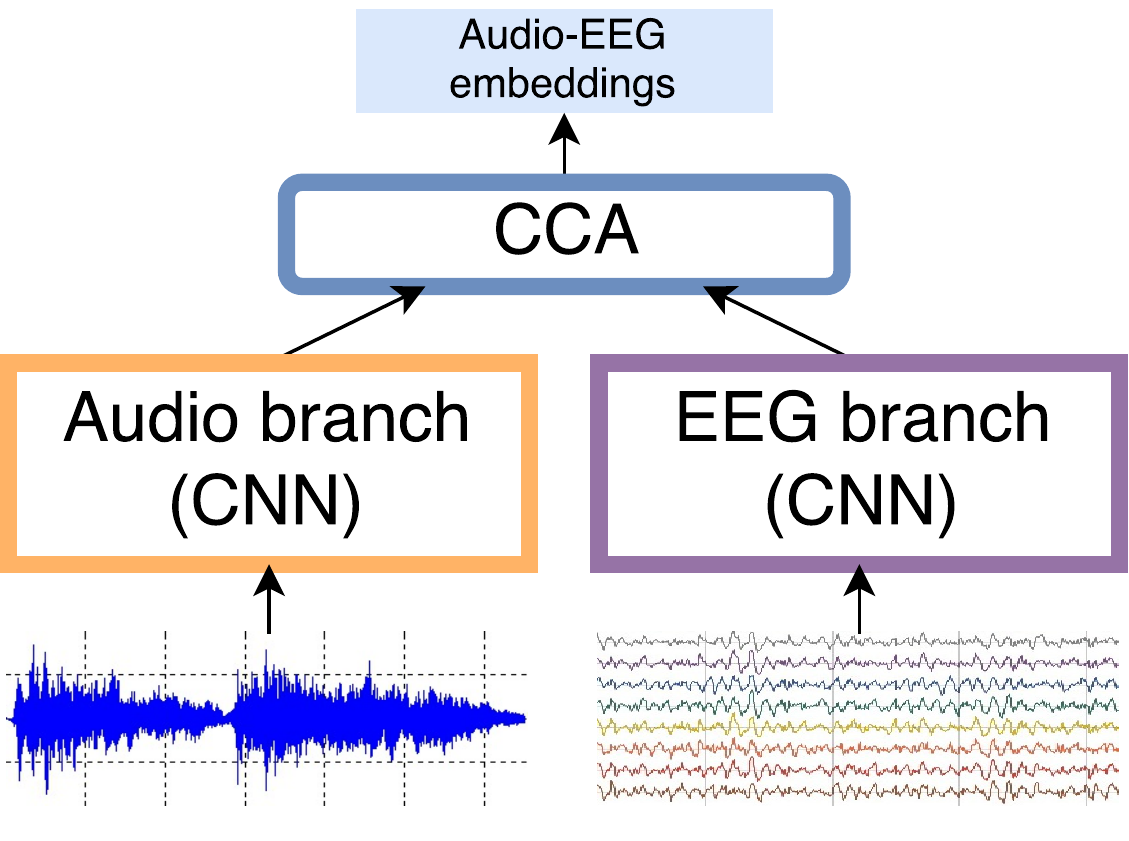}
\caption{High-level deep audio-\ac{EEG} model architecture.}
\label{fig:model}
\end{center}
\end{figure}

Following the success of sample-level \acp{CNN} in music audio modeling \cite{lee2017a}, we propose a novel fully end-to-end architecture for both views/branches of our model: audio and \ac{EEG}. It takes, as input, 1.5s signal chunks of 22050Hz-sampled mono audio and 250Hz-sampled 16-channel \acp{EEG} and outputs embeddings that are maximally correlated through their \ac{CCA} projections. We use 1D convolutional layers with ReLu non-linearities, followed by maxpooling layers. We also use batch normalization layers before each convolutional layer \cite{ioffe2015}. Window sizes were chosen so that the remainder of the integer division between the size of the input stream with the size of the output stream is 0. We refer to a convolutional layer with filter width $x$, stride length $y$, and $z$ channels as conv-$x$-$y$-$z$ and a maxpool layer with window and stride length of $x$ as mp-$x$. The audio branch is composed of the following sequence of layers: conv-3-3-128, conv-3-1-128, mp-3, conv-3-1-256, mp-3, conv-5-1-256, mp-5, conv-5-1-512, mp-5, conv-7-1-512, mp-7, conv-7-1-1024, mp-7, conv-1-1-128. The \ac{EEG} branch is: conv-3-3-128, conv-5-1-256, mp-5, conv-5-1-512, mp-5, conv-5-1-1024, mp-5, conv-1-1-128. Figure \ref{fig:model} illustrates the high-level architecture of our model.

\section{\ac{EEG} Dataset Collection}
\label{sec:dataset}

The \ac{EEG} data used in these experiments consist of two out of three subsets belonging to the same dataset, whose collection process is described in this section. All of the 18 subjects listened to 60 music segments and 2 baseline segments (noise and silence) selected by us for further research, in a randomized order. Then, each subject listened to 2 self-chosen full songs in a fixed order. Segments and full songs were separated by a 5 seconds silence interval. Each listening session took place in a quiet room, with dim light and a comfortable armchair. The subjects were asked to sit and find a relaxed position while the setup was being prepared. Then, the electrodes were placed and the subjects were asked to close their eyes and to move as little as possible, in order to avoid \ac{EOG} and \ac{EMG} artifacts. The headphones were placed and the listening session started when the subjects signaled they were ready. Subjects were informed of this setup beforehand, in order to avoid surprising them. We detail the selections for each subset below.

The first subset was built on top of a subset of a \ac{MER} dataset \cite{eerola2011}. This dataset consists of continuous clips (11.13 to 18.08 seconds, average 15.13 seconds) that were chosen in terms of dimensional and discrete emotion models. This subset consists of 60 clips but it is not used in this paper.   

The second and third subsets consist of the 2 self-chosen songs, selected according to the following criteria: one favorite song and one song that the subject does not like or does not appreciate as much, as long as that song belongs to the same artist and album as the first. The favorite song was listened to before the the second one. We use the union of both subsets (36 audio-\ac{EEG} pairs) in the experiments of this paper.

To record the \acp{EEG}, we used the OpenBCI 32bit Board with the OpenBCI Daisy Module, which provide 16 channels and up to 16kHz sampling rate. We used the default 250Hz sampling rate. The 16 electrodes were placed according to the Extended International 10-20 system on three regions of interest: frontal, central, and parietal. The locations were chosen based on the results obtained in previous studies described in Section \ref{sec:eeg}. For the frontal region of we used the Fp1, Fpz, Fp2, F7, F3, Fz, F4, and F8 locations; for the central region we used the C3, Cz, and C4 locations; and for the parietal region we used the P7, P3, Pz, P4, and P8 locations.


\section{Experimental Setup}
\label{sec:experiments}

We evaluate the semantics learned by our proposed model in a transfer learning context through a music cross-modal audio-lyrics retrieval task, using an independent dataset and model \cite{yu2017}. We compare the instance- and class-based \ac{MRR} performance of the embeddings produced by our model against a feature set available for crawling from Spotify and also against state-of-the-art embeddings. Instance-based \ac{MRR} considers that only the corresponding cross-modal object is considered as relevant, whereas in class-based \ac{MRR} any cross-modal object of the same class is considered a relevant object for retrieval. Note that we first train our proposed model with the \ac{EEG} dataset and then use this trained model as an audio feature extractor for the independent audio-lyrics dataset for performing cross-modal retrieval. The next sections present details of these experiments.

\subsection{Preprocessing}

We applied some preprocessing on the \ac{EEG} signals, namely, we remove power supply noise as well as \ac{DC} offset, with a $>$ 0.5Hz bandpass filter and a 50Hz notch filter, respectively. We attempt to perform \ac{WAR} by decomposing the signal into wavelets and then, for each wavelet, independently, removing coefficients that deviate from the mean value more than a specific multiplier (5 in our experiments) of the standard deviation and, finally, reconstruct the signals with the modified wavelets. We also use a technique called \ac{WSD} in order to remove \ac{EEG} recording noise \cite{saavedra2013}, that removes coefficients in the wavelet domain when all channels are not correlated enough, i.e., below a threshold between 0 and 1 (0.5 in our experiments). Furthermore, no matter how hard we try, the overall power of the \ac{EEG} recordings will differ across subjects, across stimuli for the same subject, and even across channels for the same subject and stimulus. This is due to loose contact between the electrodes and the scalp which is mainly caused by different people having different hair and also different head shapes. In order to circumvent this issue, we scale every \ac{EEG} signal between the values of -1 and 1 for each stimulus and channel, independently, after artifact removal but before \ac{WSD}. We also preprocess the audio signals by scaling them to fit between -1 and 1.

\subsection{Music Audio-Lyrics Dataset and Model}

\begin{figure}[htbp]
\begin{center}
\includegraphics[width=\columnwidth]{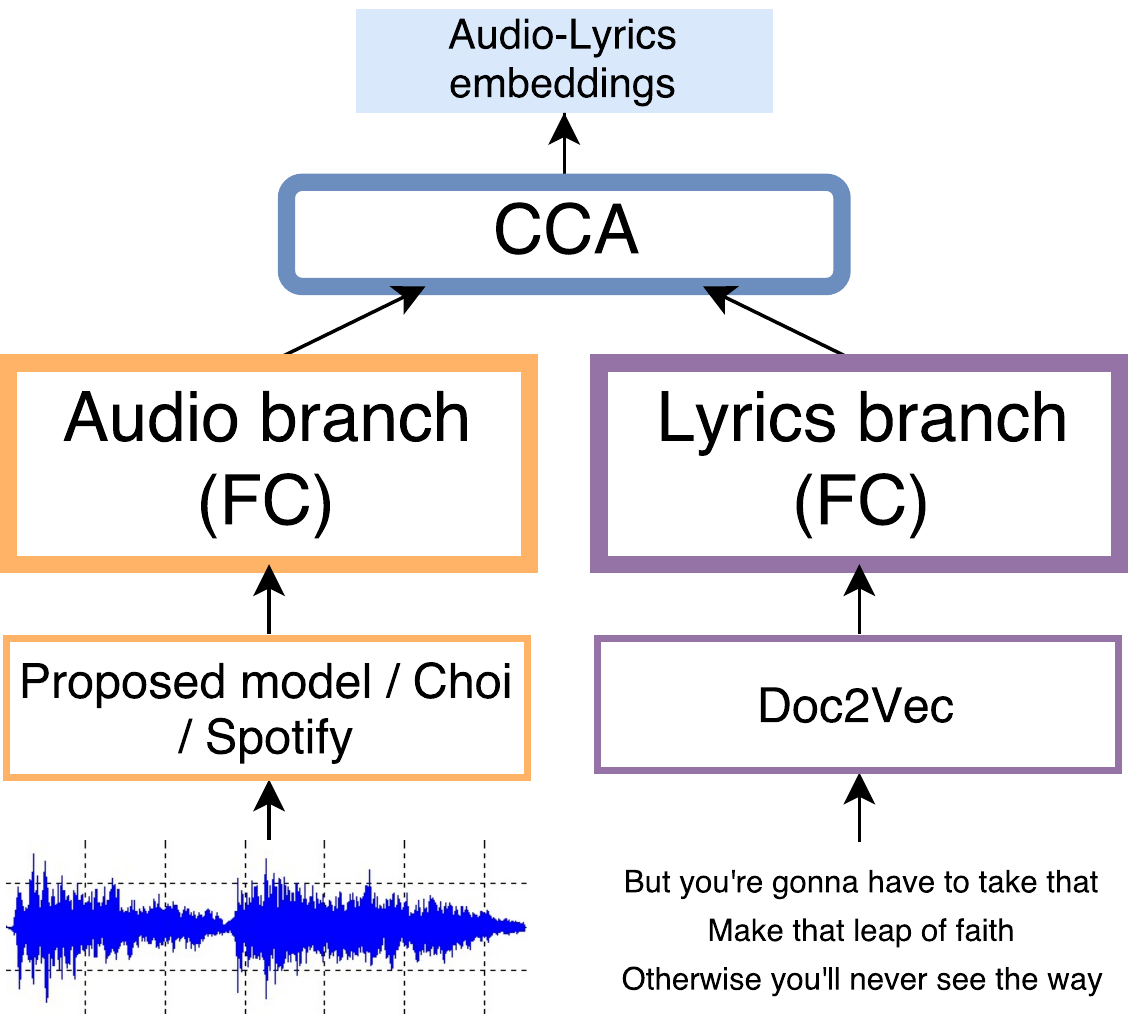}
\caption{Audio-lyrics cross-modal task setup.}
\label{fig:setup}
\end{center}
\end{figure}

We use the audio-lyrics dataset of \cite{yu2017}, implement its model, and follow its lyrics feature extraction. The \ac{NN} performing cross-modal retrieval is a 4-layer fully-connected \ac{DCCA}-based model. Layers dimensionalities for both branches are: 512, 256, 128, and 64. We use 32 canonical components. Figure \ref{fig:setup} illustrates how this model is used in the experiments.

\subsection{Baselines}

We compare the performance of our 128-dimensional embeddings against two baselines: a 65-dimensional feature vector provided by Spotify and a 160-dimensional embeddings vector from the pre-trained model of \cite{choi2017}. The Spotify set, used before in \cite{mcvicar2012}, consists of rhythmic, harmonic, high-level structure, energy, and timbre features. The pre-trained model features are computed by a \ac{CNN}-based model which was trained on supervised music tags, yet it produces embeddings that have been shown to be state-of-the-art in several tasks \cite{choi2017}. Hereby, we refer to these sets as Spotify and Choi.

\subsection{Setup}

As detailed before, our end-to-end architecture takes 1.5s of aligned audio and \acp{EEG} as input. Therefore, we segment every song and corresponding \ac{EEG} recording in 1.5s chunks for training. When predicting embeddings from this model for a new audio file, we take the average of the embeddings of all 1.5s chunks of audio as the final song-level embeddings.

We partition each dataset (audio-\ac{EEG} and audio-lyrics) into 5 balanced folds. We train our model, for 20 epochs, using 102-sized batches of size 102, 5 runs for each fold, leaving the test set out for loss function validation. This means that we have 25 different converged model instances to be used for feature extraction. Then, we run the cross-modal retrieval experiments 5 times for each feature set: our proposed embeddings, the Choi embeddings, and the Spotify features. Thus, we end up running 25$\times$5 cross-modal retrieval experiments for our proposed model. The cross-modal retrieval model is trained for 500 epochs, using batches of size 1000. We report on the average instance- and class-based \ac{MRR}.

\section{Results and Discussion}
\label{sec:results}

Table \ref{tab:mrr} shows the \ac{MRR} results. Our proposed embeddings outperform Spotify, which consists of typical handcrafted features, for this task, by 1.2 \ac{pp} for instance-based \ac{MRR} and 1.1 \ac{pp} for class-based \ac{MRR}, while performing comparably to Choi, the state-of-the-art embeddings. This is very promissing because Choi's model is trained on more than 2083 hours of music, whereas our model was trained on less than 3 hours of both music and \acp{EEG}. This also means that our model is trained faster. In fact, our model finishes training in about 20 minutes, using an NVIDIA GeForce GTX 1080 graphics card. Qualitatively, the main contribution of this approach is two-fold: (1) it provides a fine-grained and precise time alignment between the audio and \ac{EEG} regularizer data; and (2) it bypasses any fixed taxonomy selection for defining music semantics, i.e., it learns about music semantics through observation and modeling of the human brain correlates of music perception.

\begin{table}[htbp]
\normalsize
\centering
\caption{Audio-Lyrics Cross-modal retrieval results (\ac{MRR})}
\begin{tabular}{c|cc|cc}
\multirow{2}{*}{Features} & \multicolumn{2}{c|}{Instance} & \multicolumn{2}{c}{Class}\\
 & Audio & Lyrics & Audio & Lyrics\\
\hline
Spotify & 23.4\% & 23.4\% & 35.1\% & 35.1\%\\
Choi & 24.7\% & 24.8\% & 36.5\% & 36.4\%\\
Proposed model & 24.6\% & 24.6\% & 36.2\% & 36.2\%\\
\end{tabular}
\label{tab:mrr}
\end{table}

Although we already obtained good results using a simple model, they can be further improved. It is possible to learn an optimal aggregation of the embeddings of each segment using \acsp{LSTM} \cite{hochreiter1997}. Taking a personalized view for each subject is also very likely to improve the estimation of the semantic space, since having a specific set of parameters for the brain activity of each subject is, intuitively, a more realistic model. The recent success of residual learning in \acp{NN} \cite{he2016} suggests that our approach may also benefit from it. Furthermore, different loss functions for constraining the topology of the semantic space can be experimented with, including ones that impose intra-modal constraints on the embeddings to avoid destroying too much structure in each view \cite{hong2017}. When applying this framework for music discovery/recommendation, either based on audio or \ac{EEG} query, deep hashing techniques can be leveraged to design a scalable real-word system \cite{cao2017}.

\section{Conclusions and Future Work}
\label{sec:conclusions}

We proposed a novel generic framework that sets up a new approach to music semantics and a concrete architecture that implements it. We use \acp{EEG} as regularizers for learning a maximally audio-\ac{EEG} correlated space that outperforms handcrafted features and performs comparably to a state-of-the-art model that was trained with 700 times more audio data. Music embeddings can be predicted for new objects given an audio file and used for general purpose tasks, such as classification, regression, and retrieval. Future work includes a validation of these semantic spaces for music discovery as well as in other transfer learning settings. The model can be improved through several extensions, such as \acs{LSTM}, residual connections, personalized views, and other loss functions that model intra-modal constraints. Finally, it is worth studying this framework in the context of other multimedia domains.

\bibliographystyle{IEEEtran}
\bibliography{icme}


\end{document}